\documentclass{article}

\usepackage{arxiv}

\usepackage[utf8]{inputenc} 
\usepackage[T1]{fontenc}    
\usepackage{hyperref}       
\usepackage{url}            
\usepackage{booktabs}       
\usepackage{amsfonts}       
\usepackage{nicefrac}       
\usepackage{microtype}      
\usepackage{lipsum}		
\usepackage{graphicx}
\usepackage[square, sort, numbers]{natbib}
\usepackage{doi}
\usepackage{caption}
\usepackage{subcaption}

\usepackage{mathtools,upgreek,eucal,mathrsfs,enumitem,amsthm,amsmath}
\usepackage{times,epsfig,makeidx,amsfonts,amsmath,dcolumn,multirow,amssymb,colortbl,bm,url}
\usepackage{algorithm}
\usepackage[]{algorithmic}
 \usepackage[flushleft]{threeparttable}

 \DeclareMathAlphabet{\pazocal}{OMS}{zplm}{m}{n}

\newcommand{\Lb}{\pazocal{L}}

\newcommand{\btheta}{\bm{\theta}}
\newcommand{\bbeta}{\bm{\beta}}
\newcommand{\balpha}{\bm{\alpha}}

\newcommand{\bw}{{\bf w}}
\newcommand{\bx}{{\bf x}}

\newcommand{\bz}{{\bf z}}

\title{Individual frailty excess hazard models in cancer epidemiology}

\author{ 
	\href{https://orcid.org/0000-0001-7183-8407}{\includegraphics[scale=0.06]{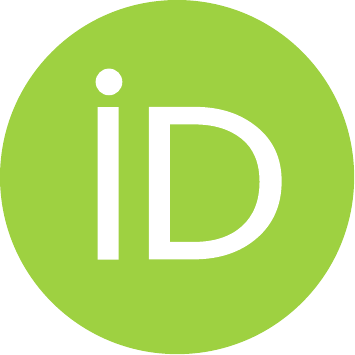}\hspace{1mm}Francisco Javier Rubio} \\
	Department of Statistical Science\\
	University College London \\
	London, UK\\
	\texttt{f.j.rubio@ucl.ac.uk}
	\And	
	\href{https://orcid.org/0000-0001-5395-1422}{\includegraphics[scale=0.06]{orcid.pdf}\hspace{1mm}Hein Putter} \\
Department of Biomedical Data Science\\
Leiden University Medical Center \\
Leiden, The Netherlands\\
	\texttt{h.putter@lumc.nl}  
\And	
	\href{https://orcid.org/0000-0003-1410-5172}{\includegraphics[scale=0.06]{orcid.pdf}\hspace{1mm}Aurelien Belot} \\
Inequalities in Cancer Outcomes Network\\
 Department of Non-Communicable Disease Epidemiology \\
	London, UK\\
	\texttt{aurelien.belot@lshtm.ac.uk} 
	}

\renewcommand{\shorttitle}{XXXX}

\hypersetup{
pdftitle={XXX},
pdfsubject={stat.ME, stat.AP},
pdfauthor={XXX, Rubio},
}

\begin{document}
\maketitle

\begin{abstract}
Unobserved individual heterogeneity is a common challenge in population cancer survival studies. This heterogeneity is usually associated with the combination of model misspecification and the failure to record truly relevant variables. We investigate the effects of unobserved individual heterogeneity in the context of excess hazard models, one of the main tools in cancer epidemiology. We propose an individual excess hazard frailty model to account for individual heterogeneity. This represents an extension of frailty modelling to the relative survival framework. In order to facilitate the inference on the parameters of the proposed model, we select frailty distributions which produce closed-form expressions of the marginal hazard and survival functions. The resulting model allows for an intuitive interpretation, in which the frailties induce a selection of the healthier individuals among survivors. We model the excess hazard using a flexible parametric model with a general hazard structure which facilitates the inclusion of time-dependent effects. We illustrate the performance of the proposed methodology through a simulation study. We present a real-data example using data from lung cancer patients diagnosed in England, and discuss the impact of not accounting for unobserved heterogeneity on the estimation of net survival. The methodology is implemented in the R package {\tt IFNS}.
\end{abstract}


\section{Introduction}
Cancer epidemiology has become one of the priorities of many countries and health organisations. The outcomes of interest in this area include the time to death since the diagnosis of cancer as well as the survival due to cancer. The latter is often interpreted as a proxy of the quality of cancer management. 
There exist different ways for analysing times to event. Briefly, the \emph{Overall Survival} framework, where the aim is to study the survival function associated with all causes of death, and the \emph{Competing Risks} framework, where the goal is to analyse survival times in the case where the death can be attributable to different causes. When the cause of death is known, the \emph{Cause-Specific} framework allows for studying cause-specific quantities, while the \emph{Relative Survival} framework allows for studying the hazard associated to cancer when the cause of death is unavailable or unreliable (\textit{e.g.}~population-based cancer registries data). The main idea behind the \emph{Relative Survival} framework is to separate the hazard associated with cancer from the hazard associated with other causes of death, without requiring to know the cause of death but using expected mortality hazard from life tables.
The overall survival framework is typically avoided in cancer epidemiology as this quantity does not represent the survival associated with cancer. Moreover, the survival of patients may be affected if the populations of interest are exposed to considerably different hazard mortality rates associated with other causes of death, thus complicating the comparison of overall survival functions associated with different populations.
Although the cause-specific framework is a useful framework for the analysis of cancer survival, the cause of death (\textit{e.g.}~from death certificates) is not available or unreliable in most countries, thus limiting its usability in population cancer epidemiology. For this reason, the relative survival framework has become the most popular tool for international comparisons of cancer survival \citep{allemani:2018}. The relative survival framework assumes that the individual hazard function can be decomposed as the hazard associated with other causes of death plus the ``excess'' hazard associated with cancer: 
\begin{eqnarray}
h(t;\bx) = h_{P}(\text{age}+t; \text{year}+t, \bz) + h_E(t;\bx),
\label{eq:exhaz}
\end{eqnarray}
where $t>0$ is the time-to-event measured since diagnosis (typically in years), $h_{P}(\text{age}+t; \text{year}+t,\bz) $ is the expected population mortality hazard obtained from a life table based on the characteristics $\bz$, ``$\text{age}$'' represents the age at diagnosis of cancer and ``$\text{year}$'' is the year of diagnosis (so $\text{age}+t$ is the age at time $t$, and $\text{year}+t$ is the year at time $t$). The expected population mortality hazard $h_{P}(\text{age}+t; \text{year}+t,\bz) $ is considered known, and the excess hazard associated with cancer, $h_E(t;\bx)$, is the quantity of interest and is estimated according to the available patient characteristics $\bx$. Typically, $\bz \subseteq \bx$ . A related quantity of interest is the \textit{net survival}, $S_N(t;\bx) = \exp\{-\int_0^{t} h_E(r;\bx)dr\}$, which is the survival function associated with the excess hazard function $h_E(t;\bx)$ for a patient with characteristics $\bx$ (see Perme et al\cite{perme:2012} and Rubio et al\cite{rubio:2019B} for a discussion on these points, and Belot et al\cite{belot:2019} for an overview of other measures used in cancer survival analysis). Policy-making and international comparisons of cancer management are often based on the net survival associated with the entire population or subgroups of interest, which is calculated as the marginal or average net survival
 $$S_N(t) = \int S_N(t;\bx)\mathrm{d} \varphi(\bx),$$
 where $\varphi$ is the distribution function of covariates $\bx$. In population studies, where the population of interest has covariates $\bx_1,\dots,\bx_m$, this quantity is usually calculated as 
$$S_N(t) = \sum_{i=1}^m S_N(t;\bx_i).$$
Several parametric \citep{lambert:2009,charvat:2016,rubio:2019S,fauvernier:2019,quaresma:2019,eletti:2022}, semi-parametric \citep{sasieni:1996}, and nonparametric \citep{perme:2012} methods have been proposed in the literature for estimating the excess hazard and the net survival. 

One of the challenges in estimating the excess hazard function is that not all of the relevant covariates may be available at the population level. 
Unobserved individual heterogeneity (UIH) refers to the situation where important covariates are not recorded, or unavailable in the sample, potentially combined with model misspecification. The effects of not accounting for UIH can be substantial. For instance, in the context of overall survival analysis, Aalen\cite{aalen:1994} and Aalen et al\cite{aalen:2014} show that neglecting UIH induces a bias on the estimation of the parameters, as well as affecting the interpretation of some epidemiological measures such as the hazard function and incidence rates. Similar effects are observed by UIH produced by model misspecification \citep{henderson:1999} or missing covariates \citep{keiding:1997}. In addition, UIH can also affect model selection \citep{rossell:2019} and the estimation of causal effects  \citep{stensrud:2017}, emphasising the importance of accounting for it. 

Individual frailty models represent a tractable option for accounting for UIH  \citep{vaupel:1979,hougaard:1995,duchateau:2007,aalen:2008,wienke:2010,hanagal:2019,balan:2019,balan:2020}. The idea behind frailty models consists of multiplying the individual hazard function by a random correction that follows a parametric distribution $G$ (see section \ref{sec:frailty} for more details on this point). This allows for accounting for non-specific departures from the original model (without a frailty term), in the sense that these departures may be a result of model misspecification or unavailable relevant covariates. 
In the relative survival framework,  Zahl \cite{zahl:1997} investigated the use of correlated frailty models, which include two random frailties that affect the hazard function associated with other causes of death and the excess hazard, separately. However, they encountered a number of inferential issues as his proposal added five parameters which cannot be simultaneously estimated, unless an arbitrary restriction of the parameter space is introduced. Indeed, jointly modelling the two competing risks (associated to other causes and cancer) is a challenging problem that may lead to non-identifiable models,\citep{tsiatis:1975} such as that proposed by Zahl \citep{zahl:1997}. 
Goeman et al.\cite{goeman:2004} studied the use of frailties in a combination of concurrent and excess hazard models. This is, assuming a further decomposition of the excess hazard into a hazard associated with a concurrent disease (assumed to be known) and the hazard associated with cancer. They assumed that the frailty term affects both the concurrent disease hazard and the excess hazard equally, and focused on modelling the excess hazard using a proportional hazards model with a piecewise baseline hazard and without time dependent effects. It is important to notice that assuming a simple model (\emph{e.g.}~without time-dependent effects) may induce a variance inflation of the frailty term, as this term captures UIH due to missing covariates or model misspecification.
Rancoita et al.\cite{rancoita:2017} investigated the use of shared frailty models, which assume that the random frailty simultaneously affects the hazard associated with other causes and the excess hazard. 
Rubio et al.\cite{rubio:2019B} studied the case where there is a potential mismatch in the estimation of the hazard associated with other causes, as a consequence of using an insufficiently stratified life table. They proposed correcting this mismatch using a random frailty affecting only the hazard associated with other causes. To our knowledge, the specific case of frailty models that can account for UIH on the excess hazard model has not been studied in the literature, neither the consequences of UIH in the estimation of net survival, which motivates our study.

In this paper, we propose a frailty excess hazard model which can account for potential UIH on the excess hazard function. This is, we assume that the UIH affects the excess hazard, either as a consequence of missing covariates or model misspecification. In order to ameliorate concerns about model misspecification, we model the excess hazard using flexible parametric models with a rich hazard structure \citep{rubio:2019S}, which can incorporate covariates that affect the time scale as well as covariates that act on the hazard scale. Such models include popular hazard structures (such as proportional hazards and accelerated failure time) as particular cases. We derive some properties of this model, and characterise the marginal hazard and survival functions for several choices of the frailty distribution. We provide intuitive interpretations of these functions, and discuss identifiability of the proposed frailty model. The resulting models are available in closed form, allowing for a tractable implementation and estimation of the parameters using maximum likelihood methods.

The paper is organised as follows. In section \ref{sec:frailty}, we present the frailty model and derive the marginal hazard and survival functions, which are required for writing down the likelihood function, and discuss the interpretation of these quantities. We consider several frailty distributions that lead to closed-form expressions of the marginal hazard and survival functions. In section \ref{sec:FAGH}, we present the flexible parametric models for the excess hazard, and discuss interpretation of the parameters. In section \ref{sec:inference}, we present the expression of the likelihood function and discuss point and interval estimation of the parameters. We discuss tools for selecting the models with and without frailties. Section \ref{sec:simulation} presents a simulation study that illustrates the performance of the proposed model in scenarios of practical interest, and illustrates the ability of the individual frailty model to recover the marginal net survival in the presence of UIH (due to the omission of one covariate). Section \ref{sec:application} presents an application using data from patients diagnosed with lung cancer in England. Section \ref{sec:discussion} concludes with a summary of our results and a general discussion about the use of frailty models in the relative survival framework. The methodology is implemented in the R package {\tt IFNS} available at \url{https://github.com/FJRubio67/IFNS}.

\section{Excess hazard frailty model}\label{sec:frailty}
In this section, we consider a frailty model based on the hazard decomposition \eqref{eq:exhaz}. This model represents an extension of frailty models \citep{balan:2020} to the relative survival framework. We calculate the marginal hazard and survival functions, which will later be used to construct the likelihood function, and discuss the interpretation of the resulting expressions. We also discuss several choices of the frailty distribution that lead to closed-form expressions of the hazard and survival functions.

Consider the following frailty model, where we include a frailty (random effect), $\lambda \sim G$, which multiplies the excess hazard function in \eqref{eq:exhaz} and is allowed to vary across individuals. More specifically, consider the conditional hazard model
\begin{eqnarray}\label{eq:GencondHazard}
\tilde{h}(t\mid \lambda; \bx) &=& h_P(\text{age}+t;\text{year}+t,\bz) + \lambda h_E(t;\bx),\\
\lambda &\sim& G.\nonumber
\end{eqnarray}
where $G$ is an absolutely continuous cumulative distribution function, with support on ${\mathbb R}_+$ and unit mean. In the Supplementary material, we present additional details about the identifiability of the proposed model under the hazard structure proposed in Section \ref{sec:FAGH}, which is guaranteed under the assumption of unit mean of the frailty distribution. The conditional survival function is 
\begin{eqnarray}
\tilde{S}(t\mid \lambda; \bx) &=& \exp\{- [H_P(\text{age}+t;\text{year}+t,\bz)-H_P(\text{age};\text{year},\bz)]\}\exp\left[-\lambda H_E(t;\bx)\right],
\label{eq:GencondSurvival}
\end{eqnarray}
where $H_P(\cdot)$ and $H_E(\cdot)$ represent the cumulative hazards associated with $h_P(\cdot)$ and $h_E(\cdot)$, respectively. Then, after integrating out the frailty $\lambda$ with respect to the distribution $G$ (see Supplementary material), the subgroup-specific marginal survival function, for a specific vector of covariates $\bx$, can be written as
\begin{eqnarray}
\tilde{S}(t;\bx) = \exp\{-H_P(\text{age}+t;\text{year}+t,\bz) + H_P(\text{age};\text{year},\bz)\}\Lb_G\{H_E(t;\bx)\},
\label{eq:MarginalSurvival}
\end{eqnarray}
where $\Lb_G\{s\} = \int_0^{\infty} e^{-s  r}dG(r)$ denotes the Laplace transform of $G$ evaluated at $s$.  Consequently, the marginal hazard function is
\begin{eqnarray}
\tilde{h}(t) &=& -\dfrac{d}{dt} \log \tilde{S}(t;\bx) \nonumber\\
&=& h_P(\text{age}+t;\text{year}+t,\bz) - \dfrac{\Lb_G'\{H_E(t;\bx)\}}{ \Lb_G\{H_E(t;\bx)\} } h_E(t;\bx)\nonumber\\
&=& h_P(\text{age}+t;\text{year}+t,\bz) + E[\lambda \mid T_o \geq t]  h_E(t;\bx),
\label{eq:MarginalHazard}
\end{eqnarray}
where $T_ o = \min \left\{ T_P, T_C\right\}$ is the observed survival time, $T_ P$ is the time to death from other causes, and $T_ C$ is the time to death from cancer, and $\Lb_G'\{ u \} = \dfrac{\partial}{\partial z} \Lb_G\{ z \} \big\vert_{z=u}$. 
A detailed derivation of the last equality is presented in the Supplementary material. Equation  \eqref{eq:MarginalHazard} reveals that the effect of the frailty on the marginal excess hazard is time-dependent and also depends on the choice of the frailty distribution $G$. This implies that the value of the marginal excess hazard, for different values of $t$, may differ for different choices of the frailty distribution (in particular, for larger values of the variance of the frailty). This becomes more evident by looking at the resulting expression of the Laplace transforms associated with different distributions (\textit{e.g.}~gamma and Inverse Gaussian, shown in the Supplementary material). Moreover, this allows us to interpret frailty excess hazard models in a similar way as in Balan et al\cite{balan:2020}, where the frailty is interpreted, at a marginal level, as an element inducing a selection of healthier individuals among cancer patients. Moreover, this factor also accounts for unobserved heterogeneity and departures from the fitted model. 
Next, we consider a specific choice for the distribution $G$: a gamma distribution. This choice allows for obtaining a closed-form expression of the marginal survival function, in addition to its appealing flexibility and interpretability of parameters. In the Supplementary material, we also present the Laplace transforms associated with the Inverse Gaussian and Power Variance Function (PVF) frailties. The PVF is a more general distribution with closed form Laplace transform, but with one additional parameter \citep{aalen:2008}. The gamma distribution is a limit case of the PVF. Other three-parameter frailty distributions leading to closed-form Laplace transforms are the compound Poisson distribution \citep{aalen:1992} and the stable distribution \cite{hougaard:1986}, which include the gamma, Inverse Gaussian, among other distributions, as particular cases. Other frailty distributions, which require numerical integration to calculate the marginal hazard and survival functions, are reviewed in Rondeau et al\cite{rondeau:2012}.

\subsubsection*{Gamma Frailty}
Consider the conditional hazard model \eqref{eq:GencondHazard} and suppose that $\lambda \sim \text{Ga}(\mu,b)$, where $\text{Ga}(\mu,b)$ denotes a gamma distribution with mean parameter $\mu > 0$, scale parameter $b>0$, and probability density function $g(r;\mu,b) = \dfrac{r^{\frac{\mu}{b}-1}}{\Gamma\left(\frac{\mu}{b}\right)b^{\frac{\mu}{b}}}\exp\left(-\dfrac{r}{b}\right)$. Under the constraint that the mean parameter $\mu$ is set to 1, the scale parameter $b$ of the frailty represents the variance. With this assumption ($\mu=1$) it follows that the subgroup-specific marginal frailty survival function is given by
\begin{eqnarray}
\tilde{S}(t;\bx) &=& \dfrac{\exp\left\{- \left[H_P(\text{age}+t;\text{year}+t,\bz)-H_P(\text{age};\text{year},\bz)\right]\right\} }{\left\{1+bH_E(t;\bx) \right\}^{\frac{1}{b}}}.
\label{eq:OverSurv}
\end{eqnarray}
The subgroup-specific marginal frailty net survival function is
\begin{eqnarray}
\tilde{S}_N(t;\bx) &=& \dfrac{1 }{\left\{1+bH_E(t;\bx) \right\}^{\frac{1}{b}}}.
\label{eq:InvMNS}
\end{eqnarray}
The subgroup-specific marginal frailty net survival \eqref{eq:InvMNS} is the net survival marginalised with respect to the frailty distribution but conditional on the covariates $\bx$. This is a corrected version of the classical net survival $S_N(t;\bx)=\exp\left[-H_E(t;\bx)\right]$, and we can see that $\lim_{b\to 0}\tilde{S}_N(t;\bx) = S_N(t;\bx)$.

The subgroup-specific marginal frailty hazard function is given by
\begin{eqnarray}
\tilde{h}(t;\bx) =  h_P(\text{age}+t;\text{year}+t,\bz) +  \dfrac{ \, h_E(t;\bx)}{1+b H_E(t;\bx)}.
\label{eq:OverHazard}
\end{eqnarray}

Expression \eqref{eq:OverHazard} provides a nice interpretation of our approach since the observed hazard can be seen as a model with a time-dependent weight function $\omega(t;\bx,b)=\dfrac{1 }{1+b H_E(t;\bx)}$ on the excess hazard, which provides a functional form that involves the scale or spread of the correction due to unobserved heterogeneity. Moreover, $\omega(t;\bx,b)$ is a decreasing function of $H_E(t;\bx)$ and $b$. Thus, an increase in these quantities induces a more pronounced frailty effect (as $0 < \omega(t;\bx,b) \leq 1$, and $\omega(t;\bx,b) = 1$ represents no frailty effect). Thus, the correction induced by the weight function $\omega(t;\bx,b)$ can be interpreted as a reduction in the level of the excess hazard $h_E(t;\bx)$, due to frailties associated with unobserved heterogeneity for specific values of the parameters. However, it is important to notice that this \emph{does not} mean that the frailty correction always shrinks the excess hazard $h_E(\cdot)$, as the model parameters are estimated \emph{jointly} with the scale parameter $b$ in the weight function. Indeed, the excess hazard in the models with and without frailty are not directly comparable. This effect will be illustrated in the simulation study and the real-data application, where we observe that the estimates of the parameters of the models with and without frailty do not necessarily coincide when there is UIH.

\section{Flexible parametric models for the excess hazard}\label{sec:FAGH}
To model the excess hazard, we consider the flexible parametric general hazard (GH) structure \cite{chen:2001}:
\begin{eqnarray}
h_E(t;\bx,\balpha,\bbeta,\btheta) = h_0\left(t \exp\{\bw^{\top}\balpha\} ; \btheta \right)\exp\left\{\bx^{\top}\bbeta\right\}.
\label{eq:HAGH}
\end{eqnarray}
where $\bx\in {\mathbb R}^{p}$, ${\bw}\in{\mathbb R}^{p_t}$, $\balpha\in{\mathbb R}^{p_t}$ and $\bbeta \in {\mathbb R}^{p}$. Typically ${\bw} \subseteq \bx$. $h_0(\cdot ;\btheta)$ represents a parametric baseline hazard function with parameters $\btheta \in \Theta$.
This is a rich hazard structure which contains, as particular cases, the Proportional Hazards (PH) model ($\balpha=0$), the Accelerated Hazards (AH) model ($\bbeta=0$), and the Accelerated Failure Time (AFT) model ($\balpha = \bbeta$, $\bw =\bx$). See \cite{rubio:2019S} for an extensive discussion on the GH structure. This structure allows for the inclusion of time-scale effects (through ${\bw}$) and effects that act at the hazard level ($\bx$). Allowing for a flexible excess hazard model helps reducing UIH associated with model misspecification, which in turn is useful to focus our attention on capturing the effect of potentially missing covariates. These points have been discussed in the overall survival framework in Gasparini et al\cite{gasparini:2019}. The corresponding cumulative hazard function is
\begin{eqnarray}
H_E(t;\bx,\balpha,\bbeta,\btheta) = H_0\left(t \exp\{{\bw}^{\top}\balpha\} ; \btheta \right)\exp\left\{\bx^{\top}\bbeta - {\bw}^{\top}\balpha\right\}.
\label{eq:CHAGH}
\end{eqnarray}
The fact that the excess cumulative hazard function can be written in closed form facilitates the implementation of the likelihood function (discussed in the next section) as well as simulating survival times from this model.
More specifically, simulating a random time-to-event from the GH model \eqref{eq:HAGH} is simple using the probability integral transform (see also Rossell and Rubio\citealp{rossell:2019}):
\begin{eqnarray}
t=\dfrac{F_0^{-1} \left[1-\exp\left\{\log(1-u)\exp(\bw^{\top}\balpha-\bx^{\top}\bbeta)\right\} ; \btheta\right]}{\exp(\bw^{\top}\balpha)},
\label{eq:simt}
\end{eqnarray}
where $u\sim (0,1)$, and $F_0$ is the cumulative distribution function associated with the baseline hazard $h_0$. Analogously, simulating from the frailty model \eqref{eq:GencondHazard} can be done as follows
\begin{eqnarray*}
t=\dfrac{F_0^{-1} \left[1-\exp\left\{\log(1-u)\exp(\bw^{\top}\balpha-\bx^{\top}\bbeta)/\lambda\right\} ; \btheta\right]}{\exp(\bw^{\top}\balpha)},
\end{eqnarray*}
where $u\sim (0,1)$, and $\lambda\sim G$.

Regarding the choice of the baseline hazard, Rubio et al\cite{rubio:2019S} presents a discussion on different choices using flexible parametric distributions. The desirable properties of this hazard function are: numerical tractability and the ability to capture the basic shapes of the hazard (increasing, decreasing, unimodal, bathtub). Rubio et al\cite{rubio:2019S} mention the Exponentiated Weibull (EW) and the Generalised gamma (GG), as particular choices for the baseline hazard. Here, we consider the use of the Power Generalised Weibull (PGW) distribution \citep{bagdonavicius:2001}, but we emphasise that any other flexible baseline hazard distribution could be used instead. The PGW distribution is a flexible three-parameter distribution (with scale parameter $\sigma>0$ and two shape parameters $\nu,\gamma>0$), which can also capture the basic hazard shapes while having tractable expressions for the hazard and survival functions \cite{alvares:2021}. The probability density function, survival function, and hazard function of the PGW distribution are presented in the Supplementary material. We could also consider simpler (2-parameter) baseline hazards that can capture specific basic shapes such as the Log-Normal distribution (unimodal), log-logistic, and the gamma distribution.

\section{Inference}\label{sec:inference}
Throughout, let $(t_1,\dots,t_n)$ be the survival times associated with $n$ cancer patients in a population; $(\delta_1,\dots,\delta_n)$ be the corresponding vital status indicators (0-alive, 1-dead); $\bx_i \in {\mathbb R}^p$, $i=,\dots,n$ represent the vector of covariates available for the $i$th patient; $\text{age}_i$ be the age at diagnosis; and $\text{year}_i$ be the year of diagnosis.
The likelihood function for the individual frailty model \eqref{eq:MarginalSurvival}--\eqref{eq:MarginalHazard} is
\begin{eqnarray*}
\tilde{L}(\balpha,\bbeta,\sigma,\nu,\gamma,b) &\propto & \prod_{i=1}^n \left[ h_P(\text{age}_i+t_i;\text{year}_i+t_i,\bz_i) +  \dfrac{h_E(t_i;\bx_i,\balpha,\bbeta,\sigma,\nu,\gamma) )}{1+b H_E(t_i;\bx_i,\balpha,\bbeta,\sigma,\nu,\gamma) )}\right]^{\delta_i} 
 \times \dfrac{1}{\left\{1+bH_E(t_i;\bx_i,\balpha,\bbeta,\sigma,\nu,\gamma) \right\}^{\frac{1}{b}}}.
\end{eqnarray*}
In contrast, the likelihood function associated with the model without frailty \eqref{eq:exhaz} is
\begin{eqnarray*}
L(\balpha,\bbeta,\sigma,\nu,\gamma) &\propto & \prod_{i=1}^n \left[ h_P(\text{age}_i+t_i;\text{year}_i+t_i,\bz_i) +  {h_E(t_i;\bx_i,\balpha,\bbeta,\sigma,\nu,\gamma) }\right]^{\delta_i}
\times
 \exp{\left\{-H_E(t_i;\bx_i,\balpha,\bbeta,\sigma,\nu,\gamma) \right\}}.
\end{eqnarray*}
Point estimates of the parameters of these models will be obtained via maximum likelihood estimation. Confidence intervals will be calculated using asymptotic normal approximations, which are justified by the typically large samples in our applications in cancer epidemiology. Briefly, let $\bm{\psi}$ be the full vector of parameters for the model of interest, we consider the use of confidence intervals of the type $\widehat{\bm{\psi}} \pm  Z_{1-\frac{\tau}{2}} \operatorname{diag}\left(J^{-\frac{1}{2}}(\widehat{\bm{\psi}})\right)$, where $J(\bm{\psi}) = -\dfrac{\partial^2}{\partial \bm{\psi} \partial \bm{\psi}^T} \log {L}(\bm{\psi})$ is the negative of the Hessian matrix of the log-likelihood function under the appropriate parametrisation, and $1-\tau \in (0,1)$ is the confidence level.

We compare these two models using AIC and evaluate their performance using a simulation study. We point out that the use of the likelihood ratio test for comparing these two models is more complex as the distribution of the likelihood ratio test statistic is not necessarily asymptotically chi-square with one degree of freedom (see, Chapter 8 of \citealp{hanagal:2019}). Moreover, there exist a number of information criteria developed for mixed models that could also be used \citep{muller:2013}. In particular, in our applications the sample size is large enough that makes the AIC and the marginal AIC \citep{muller:2013} very close. A thorough exploration of the use of different information criteria is beyond the aims of this paper and we refer the reader to M{\"u}ller et al\cite{muller:2013} for a more extensive discussion on this point.

\section{Simulation Study}\label{sec:simulation}
In this section, we present a simulation study where we investigate the performance of the frailty model under several scenarios. The proposed scenarios are designed with specific objectives \cite{burton:2006}: (Aim 1) to investigate the finite sample performance of the proposed approach in different settings, 
and (Aim 2) to assess the effect of missing covariates (with heterogeneous distributions) in subgroups of the population.  

\subsection{Aim 1: finite sample performance}
\subsubsection*{Data generation and simulation design}
We investigate the impact of (i) different sample sizes and (ii) different values of the regression coefficients on the performance of our approach. The different true values of these regression coefficients correspond to 3 different scenarios (called Sc1, Sc2 or Sc3 hereafter), and for each scenario, we consider four different sample sizes ($n \in \{500, 1000, 2000, 5000\}$). For each scenario, we generate 4 covariates: a continuous covariate (say $\text{age}$) was simulated using a mixture of uniform distributions with $0.25$ probability for the range $(30, 65)$, $0.35$ probability for $(65, 75)$  and $0.40$ probability for $(75, 85)$ years old; the remaining 3 covariates (say $\text{sex}$, $X_1$ and $X_2$) are binary with equal probabilities of being 0 or 1 (\textit{\textit{i.e.}}~ $P(\text{sex}=0)=P(\text{sex}=1)=0.5$). 

The simulation of the survival times is based on the additive decomposition of the overall mortality hazard as detailed in equation \eqref{eq:exhaz}. We simulate the ``other-causes'' time-to event using the UK life tables, and the cancer event time (\textit{i.e.}~the time to event from the excess hazard) using the inverse transform method \eqref{eq:simt}, assuming model \eqref{eq:HAGH}. This is, we adopt a GH structure with a PGW baseline excess hazard, a gamma frailty with mean $\mu^*=1$ and scale $b^*$ (its value depends on the simulated scenario), and the effects of the covariates $\text{age}$, $\text{sex}$, $X_1$ and $X_2$ (section \ref{sec:FAGH} details the use of the inverse transform method with the PGW distribution). We simulate a random drop-out time assuming an exponential distribution with a given rate $r$, the value of this rate being chosen in order to generate around 5\% of random drop-out in each scenario. Finally, we assume an administrative censoring at 5 years. Both sources of censoring (\textit{i.e.} random drop-out and administrative censoring) induce between 40\% and 45\% of censoring in total and for all the scenarios. We generate and analyse $M=1000$ samples in each scenario. 

For Sc1, the true values of the baseline distribution parameters are $\sigma^*=0.75,\nu^*=1.75,\gamma^*=8$, the scale of the gamma frailty $b^*=0.5$ and the regression coefficients $\balpha^*=\bbeta^*=(1,1,1,1)$.  We emphasise that these are complex simulation scenarios where the 4 covariates are included and with both time-dependent and hazard-level effects, in combination with a flexible baseline hazard. In practice, one would typically include only some of the covariates (\textit{e.g.}~age) as a time-dependent effect, thus reducing the effective number of model parameters. The aim of Sc1 is to portray the performance of the proposed frailty model in very challenging scenarios. 

In Sc2 and Sc3, we assumed less complex true models than in Sc1, assuming only 2 covariates with time-dependent effects. Moreover, the parameters defining the distribution of the baseline hazard and the variance of the random effect were also assumed different than in Sc1, in order to cover a suitable range of possible scenarios. The true values used for Sc2 and Sc3 can be found in the tables 1 and 2, respectively, in the Supplementary material.

\subsubsection*{Analysis of the simulated data}
We analyse the data using model \eqref{eq:HAGH} for the excess mortality hazard, assuming the same parametrisation as the one used to simulate the data.
We report the mean of the estimated regression coefficients, the bias (the difference between the mean of the estimated regression coefficients and the true value), the median of the estimated regression coefficients, the empirical standard deviation, the mean (estimated) standard error, and the coverage proportions of asymptotic confidence intervals.

Our approach has been fully implemented using R software. The optimisation step was conducted using the R command `nlminb'. The standard errors of regression coefficients were derived from the Hessian matrix obtained with the command `hessian' (R package `numDeriv'), and the asymptotic 95\% confidence intervals were approximated using these standard errors. For the initial values for the optimisation step, we used the parameters estimated from a PH model with the PGW distribution but without a frailty parameter (so we set to 1 the initial value for the scale of the gamma distribution, and for the parameters corresponding to time-dependent effects we used as initial values the ones obtained from the PH model). 
The cases where the optimiser did not converge (as indicated by `nlminb') or when the command `hessian' produced `Inf' or `NaN' values were excluded: for Sc1, it represents 31, 16, 3, and 0 sets of estimates excluded among $1000$ estimates with $n=500, 1000, 2000, 5000$ observations, respectively.
For Sc2, it represents 10, 1, 0, and 0 sets of estimates excluded among 1000 estimates with $n=500, 1000, 2000, 5000$ observations, respectively, while none were excluded in Sc3.

\begin{table}[h!]
\centering
\begingroup\scriptsize
\begin{tabular}{lrrrrrrr}
  \toprule
Parameter & True & MeanMLE & Bias & MedianMLE & Coverage & Mean StdErr & EmpSD \\ 
  \midrule
  \textbf{Scenario N=500} \\
  $\sigma$ & 0.750 & 0.974 & 0.224 & 0.761 & 0.817 & 0.446 & 0.622 \\ 
  $\nu$ & 1.750 & 1.983 & 0.233 & 1.873 & 0.932 & 0.397 & 0.498 \\ 
  $\gamma$ & 8.000 & 8.690 & 0.690 & 8.609 & 0.787 & 4.097 & 4.986 \\ 
  $\alpha_1$ &   1.000 & 1.097 & 0.097 & 1.006 & 0.928 & 0.607 & 1.333 \\ 
  $\alpha_2$ &   1.000 & 0.984 & -0.016 & 0.995 & 0.914 & 0.273 & 0.639 \\ 
  $\alpha_3$ &   1.000 & 1.027 & 0.027 & 0.991 & 0.924 & 0.565 & 1.197 \\ 
  $\alpha_4$ &   1.000 & 1.010 & 0.010 & 0.986 & 0.917 & 0.608 & 1.420 \\ 
  $\beta_1$ &   1.000 & 0.940 & -0.060 & 1.016 & 0.941 & 0.384 & 0.846 \\ 
  $\beta_2$ &   1.000 & 1.007 & 0.007 & 0.996 & 0.964 & 0.165 & 0.369 \\ 
  $\beta_3$ &   1.000 & 0.989 & -0.011 & 1.009 & 0.957 & 0.359 & 0.675 \\ 
  $\beta_4$ &   1.000 & 1.003 & 0.003 & 1.014 & 0.946 & 0.378 & 0.821 \\ 
  $b$ & 0.500 & 1.231 & 0.731 & 0.549 & 0.828 & 0.810 & 1.675 \\ 
  \midrule
 \textbf{Scenario N=1000} \\
 $\sigma$   &  0.750 & 0.870 & 0.120 & 0.740 & 0.885 & 0.333 & 0.453 \\ 
 $\nu$      & 1.750 & 1.847 & 0.097 & 1.820 & 0.937 & 0.246 & 0.269 \\ 
 $\gamma$   & 8.000 & 8.364 & 0.364 & 8.246 & 0.869 & 3.065 & 3.404 \\ 
 $\alpha_1$ &  1.000 & 1.029 & 0.029 & 1.000 & 0.957 & 0.319 & 0.582 \\ 
 $\alpha_2$ &  1.000 & 1.023 & 0.023 & 1.001 & 0.955 & 0.162 & 0.373 \\ 
 $\alpha_3$ &  1.000 & 1.004 & 0.004 & 0.972 & 0.953 & 0.319 & 0.570 \\ 
 $\alpha_4$ &  1.000 & 0.969 & -0.031 & 0.989 & 0.945 & 0.326 & 0.584 \\ 
 $\beta_1$  &   1.000 & 0.994 & -0.006 & 1.008 & 0.954 & 0.189 & 0.263 \\ 
 $\beta_2$  &   1.000 & 0.984 & -0.016 & 0.995 & 0.948 & 0.090 & 0.207 \\ 
 $\beta_3$  &   1.000 & 0.997 & -0.003 & 1.005 & 0.952 & 0.191 & 0.327 \\ 
 $\beta_4$  &   1.000 & 1.007 & 0.007 & 0.999 & 0.956 & 0.194 & 0.317 \\ 
  $b$       & 0.500 & 0.836 & 0.336 & 0.541 & 0.923 & 0.616 & 1.063 \\ 
  \midrule
 \textbf{Scenario N=2000} \\
  $\sigma$ &  0.750 & 0.821 & 0.071 & 0.759 & 0.942 & 0.227 & 0.309 \\ 
   $\nu$ & 1.750 & 1.778 & 0.028 & 1.766 & 0.962 & 0.165 & 0.162 \\ 
  $\gamma$ & 8.000 & 8.061 & 0.061 & 8.010 & 0.929 & 2.170 & 2.292 \\ 
 $\alpha_1$ & 1.000 & 1.015 & 0.015 & 0.997 & 0.959 & 0.192 & 0.272 \\ 
 $\alpha_2$ & 1.000 & 1.010 & 0.010 & 1.005 & 0.954 & 0.093 & 0.154 \\ 
 $\alpha_3$ & 1.000 & 1.009 & 0.009 & 0.991 & 0.945 & 0.197 & 0.338 \\ 
 $\alpha_4$ & 1.000 & 1.009 & 0.009 & 0.995 & 0.945 & 0.193 & 0.310 \\ 
 $\beta_1$ &  1.000 & 1.003 & 0.003 & 1.008 & 0.947 & 0.113 & 0.147 \\ 
 $\beta_2$ &  1.000 & 0.994 & -0.006 & 0.998 & 0.956 & 0.048 & 0.087 \\ 
 $\beta_3$ &  1.000 & 0.996 & -0.004 & 1.004 & 0.946 & 0.116 & 0.185 \\ 
 $\beta_4$ &  1.000 & 0.997 & -0.003 & 1.003 & 0.939 & 0.114 & 0.175 \\ 
  $b$ & 0.500 & 0.660 & 0.160 & 0.550 & 0.964 & 0.409 & 0.607 \\ 
  \midrule
\textbf{Scenario N=5000} \\
  $\sigma$ &  0.750 & 0.760 & 0.010 & 0.747 & 0.950 & 0.123 & 0.125 \\ 
   $\nu$ & 1.750 & 1.764 & 0.014 & 1.760 & 0.954 & 0.102 & 0.102 \\ 
  $\gamma$ & 8.000 & 8.108 & 0.108 & 8.026 & 0.951 & 1.352 & 1.369 \\ 
 $\alpha_1$ &  1.000 & 0.991 & -0.009 & 0.987 & 0.952 & 0.109 & 0.110 \\ 
 $\alpha_2$ &  1.000 & 1.001 & 0.001 & 1.002 & 0.949 & 0.051 & 0.052 \\ 
 $\alpha_3$ &  1.000 & 0.988 & -0.012 & 0.986 & 0.948 & 0.110 & 0.109 \\ 
 $\alpha_4$ &  1.000 & 0.987 & -0.013 & 0.989 & 0.936 & 0.110 & 0.113 \\ 
 $\beta_1$ &   1.000 & 0.999 & -0.001 & 0.997 & 0.945 & 0.066 & 0.066 \\ 
 $\beta_2$ &   1.000 & 1.000 & -0.000 & 0.998 & 0.949 & 0.027 & 0.028 \\ 
 $\beta_3$ &   1.000 & 1.001 & 0.001 & 1.000 & 0.943 & 0.066 & 0.067 \\ 
 $\beta_4$ &   1.000 & 1.002 & 0.002 & 1.001 & 0.938 & 0.066 & 0.068 \\ 
  $b$ & 0.500 & 0.565 & 0.065 & 0.543 & 0.941 & 0.229 & 0.233 \\ 
   \bottomrule
\end{tabular}
\endgroup
\caption{Simulation results for Aim1, scenario 1. MeanMLE: Mean of the Maximum Likelihood Estimates; MedianMLE: median value of the Maximum Likelihood Estimates; Mean StdErr: Mean of the standard errors; EmpSD: Empirical standard deviation} 
\label{Table_Sc1}
\end{table}

\subsubsection*{Results}
The results of Sc1 are detailed in Table \ref{Table_Sc1} below, while the results for Sc2 and Sc3 are given in Tables 1 and 2 in the Supplementary material, respectively. As expected, the bias decreases and the coverage gets closer to the nominal value with increasing sample size. Using different values of regression coefficients and variance of the frailty does not alter the performance and the results observed, as shown in Tables 1 and 2 in the Supplementary material. 
We observed a different behaviour in the estimation of the regression coefficients associated with covariates compared to those of the baseline hazard parameters and the scale parameter of the frailty. 
For the regression coefficients associated with covariates, the performances are good even with a small sample size of 500 cases. For the baseline distribution parameters and the scale of the frailty, the bias is high in situations with sample size of 500 or 1000. When the sample size is larger ($2000$ or $5000$), the performances are good, with small bias and coverage near the nominal value.  

\begin{figure}[h!]
	\centering
	\includegraphics[scale = 0.3]{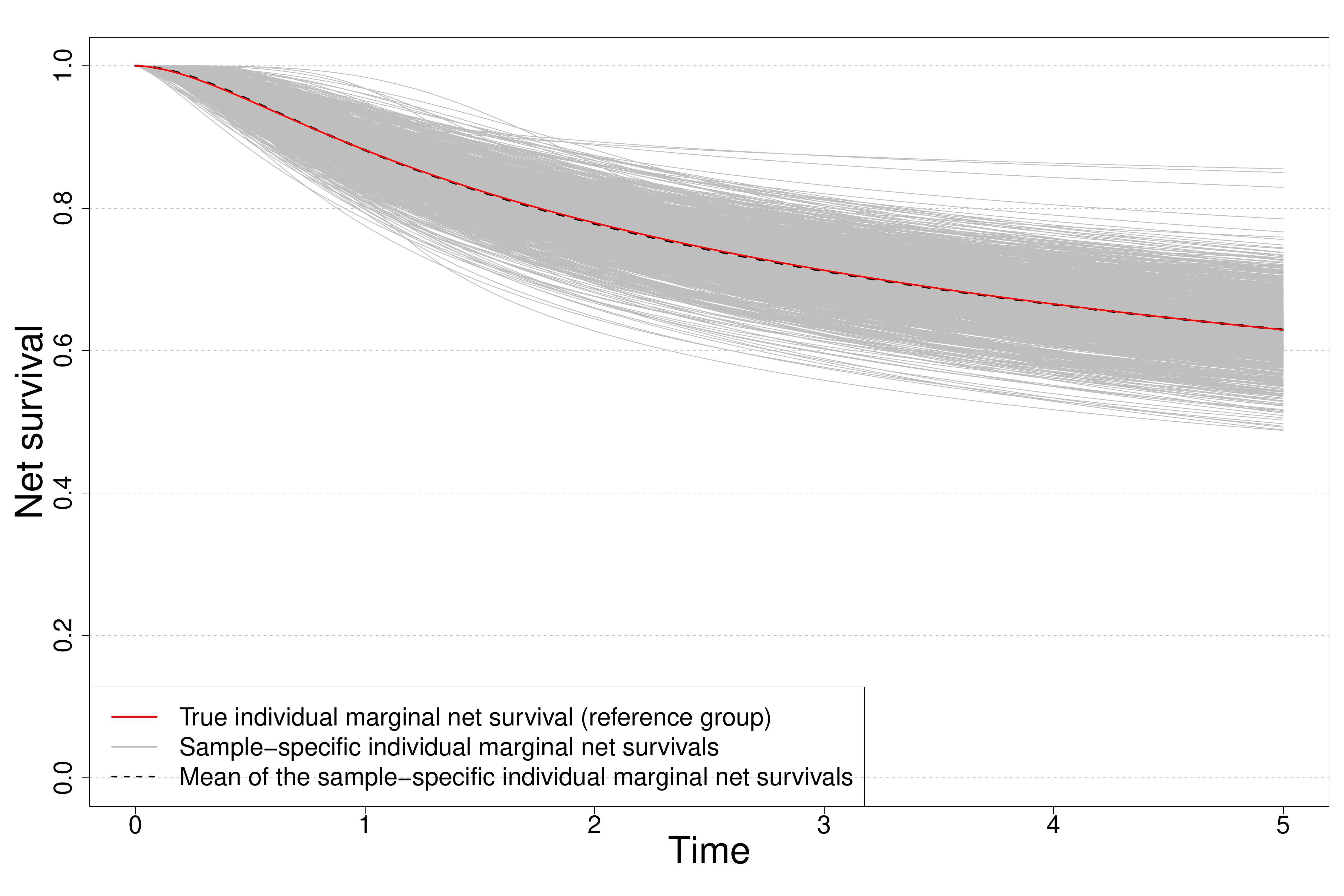}
	\caption{Scenario Sc1 with 500 observations: subgroup-specific marginal net survival for the reference group}
	\label{fig:InvMNS}
\end{figure}

For Sc1 with 500 observations (and therefore non negligible bias on the baseline distribution parameters and the scale of the frailty), we check the ability of our approach to recover the true subgroup-specific marginal net survival (\textit{i.e.} integrated over the frailty distribution but conditional on covariates, see equation \eqref{eq:InvMNS}. We could see that for the reference group (\textit{i.e.} covariates values set to 0) and despite the bias aforementioned, the true subgroup-specific marginal net survival is nicely recovered on average over the set of simulated samples (Figure \ref{fig:InvMNS}). 

\subsection{Aim 2: impact of missing covariates with heterogeneous distributions}
We now explore two simulation scenarios as follows. The overall aim is to compare the use of the model fitted to the entire population for predicting net survival for subgroups of the population, a common strategy in epidemiology, against a stratified analysis.
In the first scenario, we assume that there exist two subgroups of the population, defined by the variable $\text{sex}$, with slightly different PGW baseline hazards ($\btheta_1 = (0.5, 1.5, 5)$ for $\text{sex} = 1$, and $\btheta_2 = (0.5, 1.5, 3)$, for $\text{sex} = 0$). The variable $\text{sex}$ is binary with probability $0.6$ of being 1 and probability $0.4$ of being 0.
For each subgroup, we generate 2 covariates: a continuous covariate (say $\text{age}$) was simulated using a mixture of uniform distributions with $0.25$ probability for the range $(30, 65)$, $0.35$ probability for $(65, 75)$  and $0.40$ probability for $(75, 85)$ years old; and the variable $X_1$ is also binary, with conditional probability $0.8$ of being 1 if $\text{sex} = 1$, an conditional probability $0.4$ of being 1 if $\text{sex} = 0$ (that is, it has a different distribution for each \text{sex}). In addition, we assume that the effects of the covariates are different for each sex subgroup: for $\text{sex} = 1$, $\balpha = (0.7, 0.7, 0.5)$ and $\bbeta = (1, 0.5, 1)$; and for $\text{sex} = 0$, $\balpha = (0.7, 0.7, 0.25)$ and $\bbeta = ( 0.5, 0.5, 0.25)$. 
In the second simulation scenario, we again assume that there exist two subgroups of the population, defined by the variable $\text{sex}$, now with markedly different PGW baseline hazards ($\btheta_1 = (0.5, 1.5, 5)$ for $\text{sex} = 1$, and $\btheta_2 = (0.5, 1.5, 0.75)$ for $\text{sex} = 0$). The covariates are simulated as in the first scenario, and the true values of the regression coefficients $\balpha$ and $\bbeta$ are chosen as in the first scenario. For both scenarios, we consider sample sizes $n=500,1000,5000$ and censoring rate of approximately $65\%$.

For each scenario, we first fit the classical model without frailty \eqref{eq:exhaz} and the frailty model \eqref{eq:OverHazard} with PGW baseline hazard to each simulated data set omitting the variable $X_1$. 
We  calculate the corresponding population net survival curves by averaging the individual net survival curves, for both models, associated to each patient. 
We compare the true population net survival curve against the fitted net survival curves obtained with these models. Then, we calculate the net survival associated to the subgroups defined by $\text{sex} = 0,1$ by taking the average of individual net survival curves over the corresponding subsets of observations.
Our aim here is to compare the net survival curves associated to the entire population, and the use of the models fitted to the entire sample for net survival subgroup estimations in the presence of UIH.

In a second stage, we perform a stratified analysis where we fit the classical model without frailty \eqref{eq:exhaz} and the frailty model \eqref{eq:OverHazard} to each data \emph{subset} defined by the variable $\text{sex} = 0,1$.
Then, we calculate the net survival curves associated to the subgroups $\text{sex} = 0,1$ with the average of the individual fitted net survival curves obtained with these models.
We again compare the corresponding net survival curves against the true net survival for these subgroups. 

Results are shown in the Supplementary material (Figures 1--2). In the first scenario and for all sample sizes (Figure 1 in the Supplementary material), we notice that the average fitted population net survival curves are very close to the true population net survival. The subgroup analysis also leads to close net survival curves, with a slightly worst fit from the classical model (which remains for all sample sizes). The stratified analysis produces the best results as the average fitted net survival curves are virtually the same as the true net survival curves. In the second scenario (Figure 2 in the Supplementary material), we notice that the average fitted population net survival curves are close to the true population net survival, although a small bias is observed for both models and all sample sizes. The subgroup analysis leads to a very large bias for both average curves, and we also notice a discrepancy between the classical and frailty models. This bias is largely reduced by using a stratified analysis for all sample sizes.

\section{Real data application: lung cancer epidemiology}\label{sec:application}

We analyse a data set obtained from population-based national cancer registry of Non-Small Cell Lung Cancer (NSCLC) female patients diagnosed in 2012 in England \citep{belot2019association}. We include the following covariates:  standardised age at diagnosis (\texttt{agec}, continuous); tumour stage at diagnosis (\texttt{stage}, categorical I-IV); the presence of cardiovascular comorbidities (\texttt{CVD}, binary); the presence of Chronic Obstructive Pulmonary Disease (\texttt{COPD}, binary); and the standardised Income Domain from the 2010 England Indices of Multiple Deprivation (\texttt{IMD}, continuous), defined at the Lower Super Output Area level (as a measure of deprivation). Information on stage at diagnosis and comorbidities was obtained from linked data (Hospital Episode Statistics -HES- and the Lung Cancer Audit data). The cohort of patients was followed-up until the 31st of December 2015, at which time patients alive were right-censored. We restrict the analysis to women with no missing covariates. The resulting sample size was $n=14557$ patients with complete cases, among which $n_o=12138$ died before the 31st of December 2015. The $25\%$, $50\%$ and $75\%$ quantiles of the patients' age at diagnosis was $64.93$, $72.64$, $80.23$ while the mean was $72.00$. Among these patients, $2434 $ were diagnosed at Stage I, $1131$ at Stage II, $3241$ at Stage III, and $7751$ at Stage IV. Finally, $1954$ patients were classified with a cardiovascular comorbidity and $3260$ with a chronic obstructive pulmonary disease.

In order to evaluate the effects of UIH, we compare eight models for this data set. 
\textit{Model C1}: (classical) model without frailty with time-level effect of \texttt{agec} (${\bw}_i$), and hazard-level ($\bx_i$) effects of \texttt{agec}, \texttt{IMD}, \texttt{stage}, \texttt{CVD}, and \texttt{COPD}.
\textit{Model F1}: frailty model with time-level effect of \texttt{agec} (${\bw}_i$), and hazard-level ($\bx_i$) effects of \texttt{agec}, \texttt{IMD}, \texttt{stage}, \texttt{CVD}, and \texttt{COPD}.
\textit{Model C2}: model without frailty with time-level effect of \texttt{agec} (${\bw}_i$), and hazard-level ($\bx_i$) effects of \texttt{agec}, \texttt{IMD}, \texttt{CVD}, and \texttt{COPD}.
\textit{Model F2}: frailty model with time-level effect of \texttt{agec} (${\bw}_i$), and hazard-level ($\bx_i$) effects of \texttt{agec}, \texttt{IMD}, \texttt{CVD}, and \texttt{COPD}.
\textit{Model C3}: model without frailty with time-level effect of \texttt{agec} (${\bw}_i$), and hazard-level ($\bx_i$) effects of \texttt{agec}, \texttt{IMD}.
\textit{Model F3}: frailty model with time-level effect of \texttt{agec} (${\bw}_i$), and hazard-level ($\bx_i$) effects of \texttt{agec}, \texttt{IMD}. 
\textit{Model C4}: PGW baseline hazard without covariates. 
\textit{Model F4}: frailty model with PGW baseline hazard without covariates.

\begin{figure}[h!]
\centering
\begin{tabular}{c c}
\includegraphics[scale = 0.4]{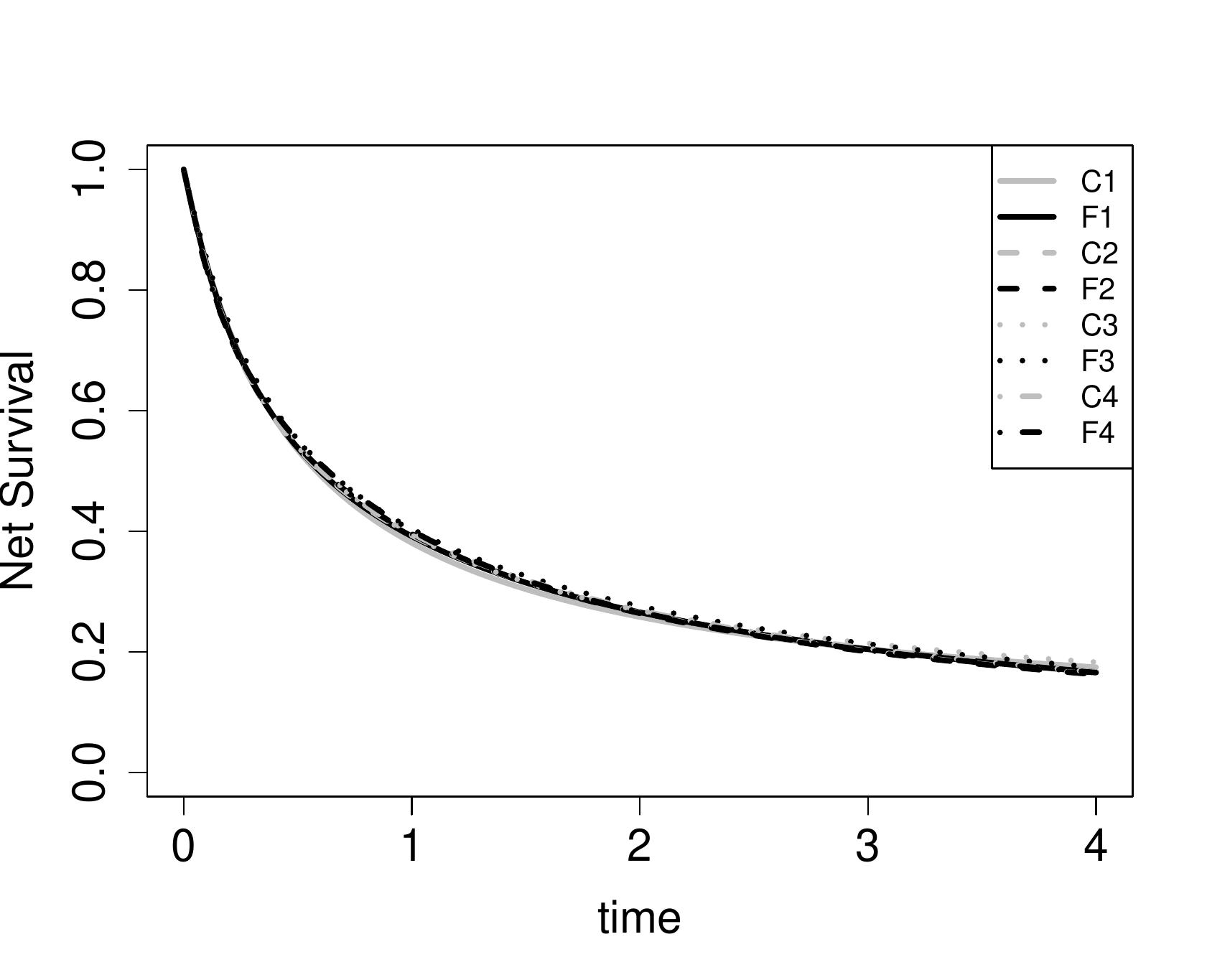} &
\includegraphics[scale = 0.4]{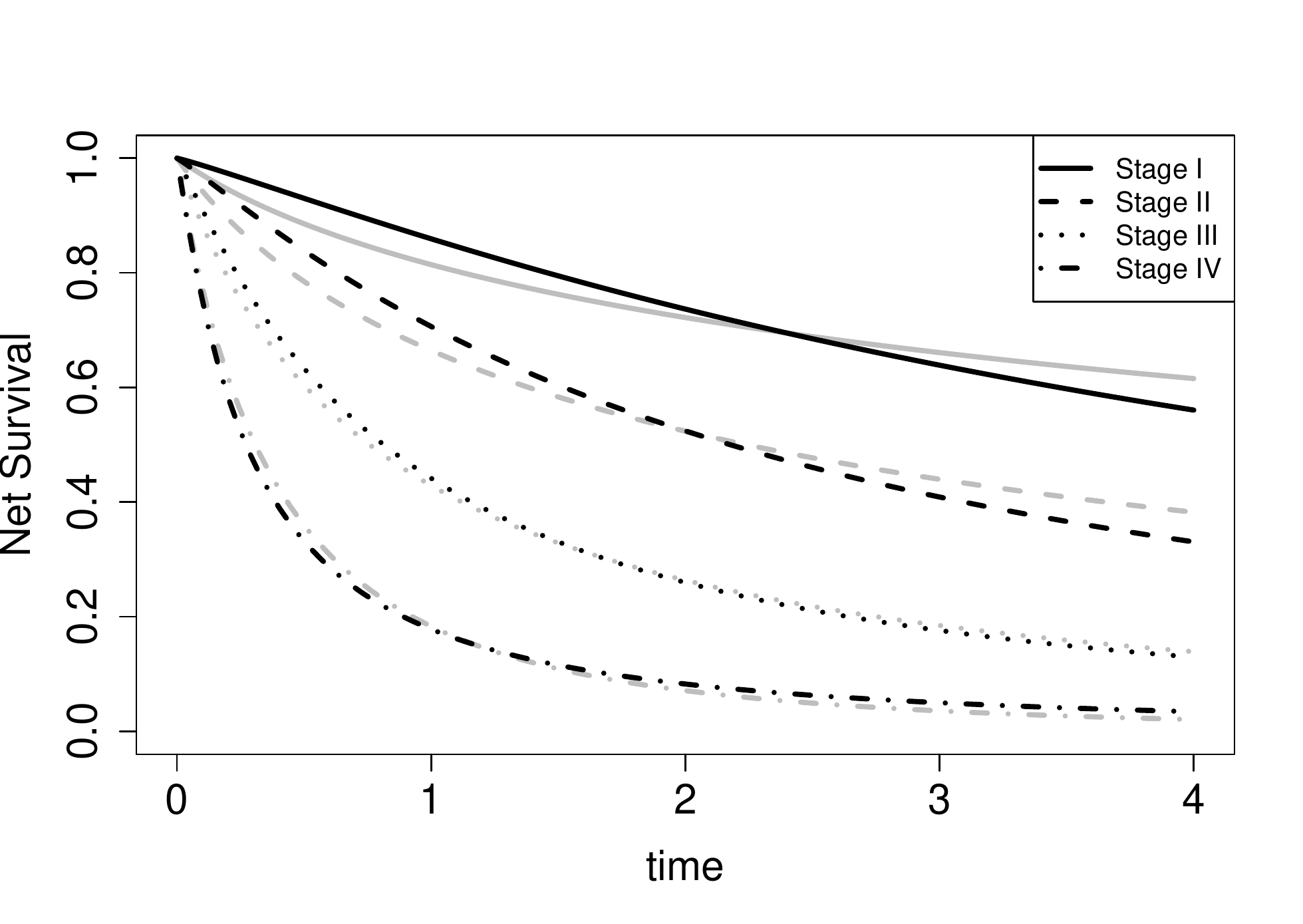} \\
(a) & (b)
\end{tabular}
\caption{Lung cancer data: (a) net survival curves for the entire population, and (b) stratified net survival curves by tumour stage at diagnosis (Model C1 - gray lines, Model F1 - black lines).}
\label{fig:netsurvlung}
\end{figure}

Figure \ref{fig:netsurvlung}a shows the net survival for the entire cohort obtained with Models C1--C4 and F1--F4. These are obtained using the formulas:
\begin{eqnarray*}
S_N(t)  &=&  \dfrac{1}{n}\sum_{i=1}^n \exp\{-H_E(t; \bx_i, \widehat{\balpha},\widehat{\bbeta},\widehat{\sigma},\widehat{\nu},\widehat{\gamma})\}, \; \text{Models C1--C4},\\
\tilde{S}_N(t)&=& \dfrac{1}{n}\sum_{i=1}^n \dfrac{1 }{\left\{1+\widehat{b} H_E(t; \bx_i, \widehat{\balpha},\widehat{\bbeta},\widehat{\sigma},\widehat{\nu},\widehat{\gamma}) \right\}^{\frac{1}{\widehat{b}}}}, \; \text{Models F1--F4}.
\end{eqnarray*}
The AIC values for Models C1--C4 and F1--F4 are shown in Table \ref{tab:MLES}. The overall best model is Model F1, clearly favouring a model with all covariates and a frailty. 
The best model among the classical models (C1--C4) was model C1, also favouring the inclusion of all covariates. Some interesting conclusions arise from this model comparison and Figure \ref{fig:netsurvlung}. Although AIC clearly favours Model F1, Figure \ref{fig:netsurvlung}a shows that the estimated population net survival curves obtained for models C1--C4 and F1--F4 are very similar. This indicates that, even though the data favours a frailty model, some estimated quantities based on models with and without frailties may be similar. However, this is far from being a general conclusion. Figure \ref{fig:netsurvlung}b shows that the net survival stratified by \texttt{stage} for models C1 and F1 exhibit marked differences for early tumour stages. These differences are masked in Figure \ref{fig:netsurvlung}a due to the distribution of the tumour stage covariate, which mostly contains late-stage patients (see the descriptive analysis).
This is, since we are interested in estimating quantities based on averages over subgroups of the population, the distribution of the corresponding covariates (patient characteristics) plays a key role.
Thus, although the estimated models are quite different (see Table \ref{tab:MLES}), the distribution of the covariates produces similar marginal net survival curves at the population level. However, this conclusion does not apply to the net survival functions stratified by \texttt{stage}. Table \ref{table:confintNS} presents confidence intervals for the net survival at times $t=1,2,3,4$ years. These confidence intervals are obtained using a Monte Carlo approximation based on the asymptotic normality of the maximum likelihood estimators (see Rubio et al\cite{rubio:2019S} for more details on this). These confidence intervals confirm that differences in the estimation of net survival are also present in interval estimates.

Another point that deserves some attention is the interpretation of parameters. From Table \ref{tab:MLES}, we can see that the estimates of the parameters of the baseline hazards markedly differ for models C1 and F1. The reason for this is that the estimated baseline hazard associated to model C1 is decreasing, while the estimated baseline hazard associated to model F1 is increasing. This explains the differences in the estimates of the parameters, which are simply reflecting this contrasting shape. However, we would like to emphasise that the baseline hazards associated to models C1 and F1 (more generally, models with and without frailty correction) are not directly comparable as (i) the estimates for model F1 correspond to a marginal model (marginalised over the frailty term), and (ii) the baseline hazard in the frailty model \eqref{eq:OverHazard} controls both the excess hazard in the numerator and the time-varying weight. Despite this difference in nature, comparing them is still useful, as observing differences between the estimates in both models is an indication of UIH as such discrepancies only appear when the variance of the frailty term is non-negligible. 

\begin{table}[h!]
\centering
\begin{tabular}{ccccccccc}
  \hline
Model & C1 & F1 & C2 & F2 & C3 & F3 & C4 & F4 \\ 
  \hline
$\widehat{\sigma}$    & 0.763 & 8.760 (7.072,10.851) & 0.109 & 0.769 & 0.107 & 0.737 & 0.099 & 0.562 \\ 
$\widehat{\nu}$          & 0.987 & 1.077 (1.046,1.108) & 1.196 & 1.273 & 1.199 & 1.288 & 1.197 & 1.407 \\ 
$\widehat{\gamma}$  & 4.980 & 0.813 (0.675,0.980) & 4.185 & 0.323 & 4.173 & 0.314 & 4.325 & 0.272 \\ 
$\texttt{agec}_t$        & 0.308 & 2.633 (1.367,3.900) & 0.271 & 0.442 & 0.272 & 0.436 &  --       & --  \\ 
$\texttt{agec}$           & 0.328 & 0.124 (0.000 0.248) & 0.331 & 0.289 & 0.340 & 0.303 &  --       & -- \\ 
$\texttt{IMD}$            & 0.530 & 0.827 (0.578,1.075) & 0.305 & 1.079 & 0.319 & 1.126 & --        & --  \\ 
$\texttt{stage2}$        & 0.706 & 0.964 (0.825,1.102) & --       & --        & --       & --        & --        & --  \\ 
$\texttt{stage3}$        & 1.460 & 2.085 (1.963,2.207) & --       & --        & --        & --        &  --      & --  \\ 
$\texttt{stage4}$        & 2.202 & 3.312 (3.169,3.456) & --       & --        & --       & --         & --       &  -- \\ 
$\texttt{CVD}$           & 0.214 & 0.352 (0.267,0.437) & 0.147 & 0.295  & --      & --         & --       & --  \\ 
$\texttt{COPD}$        & 0.142 & 0.242 (0.172,0.311) & -0.018 & -0.054  & --     & --       & --        &  -- \\ 
$\widehat{b}$            & --       & 0.764 (0.687,0.850) & --         & 4.542   & --      & 4.785 & --      & 6.836 \\ 
AIC                            & 20548.43 & \textbf{20140.06} & 26328.84 & 26311.62 & 26306.3 & 26302.23 & 
26864.78 & 26808.51 \\
   \hline
\end{tabular}
\caption{Maximum likelihood estimates and AIC for Models C1--C4 and F1--F4. $95\%$ confidence intervals are also presented for the parameters of the selected model, F1.}
\label{tab:MLES}
\end{table}

\begin{table}
\centering
\begin{tabular}{|cccccccc|}
\hline
\multicolumn{8}{|c|}{\underline{Total population by Stage}} \\
&&&&&&&\\
& & \multicolumn{3}{c}{Classical (C1)} &  \multicolumn{3}{c|}{Frailty (F1)} \\
&&&&&&&\\
Stage & year & NS & lower & upper & NS & lower & upper \\
\multirow{4}{*}{I}   & 1 & 0.814 & 0.801 & 0.827 & 0.859 & 0.850 & 0.873 \\ 
&  2 & 0.722 & 0.704 & 0.740 & 0.737 & 0.721 & 0.760 \\ 
 & 3 & 0.661 & 0.640 & 0.681 & 0.639 & 0.620 & 0.668 \\ 
 & 4 & 0.615 & 0.590 & 0.637 & 0.560 & 0.542 & 0.591 \\ 
&&&&&&&\\
\multirow{4}{*}{II}   & 1 & 0.664 & 0.642 & 0.688 & 0.706 & 0.685 & 0.732 \\ 
&  2 & 0.524 & 0.497 & 0.551 & 0.524 & 0.501 & 0.557 \\ 
 & 3 & 0.440 & 0.413 & 0.466 & 0.409 & 0.385 & 0.442 \\ 
 & 4 & 0.382 & 0.355 & 0.411 & 0.331 & 0.309 & 0.359 \\ 
\hline
\end{tabular}
\caption{Net survival (NS) at $t=1,2,3,4$ years and $95\%$ confidence intervals: Classical Model (Model C1) and Frailty Model (Model F1).}
\label{table:confintNS}
\end{table}

\subsection*{Stratified analysis}
\begin{figure}[h!]
\centering
\includegraphics[scale = 0.65]{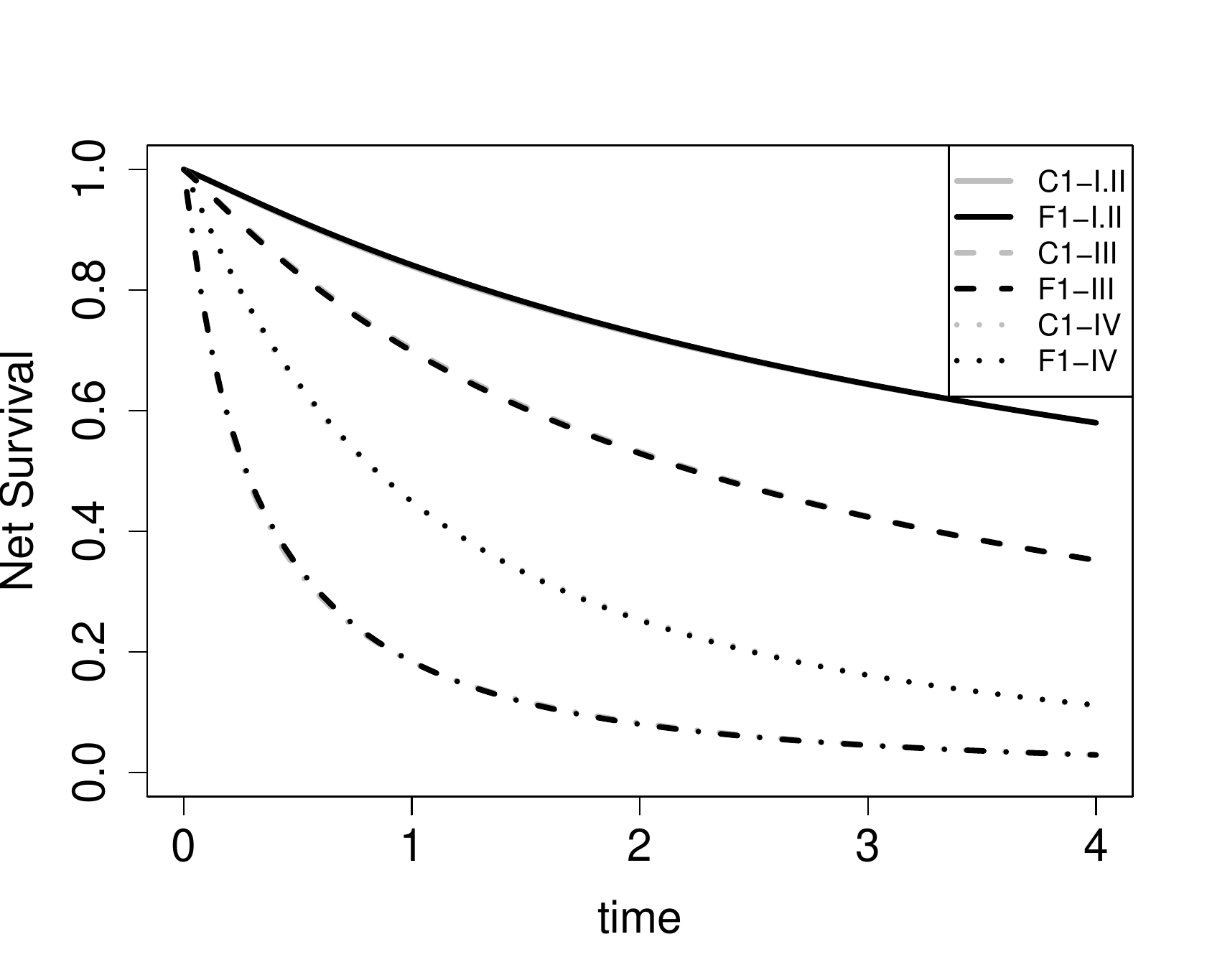} \\
\caption{Lung cancer data: net survival curves for the models fitted on the stratified data by Stage. Gray lines represent the net survival curves associated to model C1 fitted to the corresponding stratum (Stages I-II, Stage III, or Stage IV). Black lines represent the net survival curves associated to model F1 fitted to the corresponding stratum (Stages I-II, Stage III, or Stage IV).}
\label{fig:netsurvlung_strat}
\end{figure}

The differences in the net survival curves for Stages I--IV in Figure \ref{fig:netsurvlung} suggest that there is UIH, and that, potentially, the unobserved covariates may have a different conditional distribution at each of the strata. 
To explore this idea, we now consider a stratified analysis where we study patients in three different strata: Stage I-II, Stage III, and Stage IV tumours. There are clinical and biological reasons for this kind of stratification, as different tumour stages indicate the size of the tumour and spread to other organs (metastasis). 
We compare frailty and non-frailty models for each of these groups to evaluate the effects of UIH. For Stages I--II we fit the models: \textit{Model C1-I.II}: model without frailty with time-level effect of \texttt{agec} (${\bw}_i$), and hazard-level ($\bx_i$) effects of \texttt{agec}, \texttt{IMD}, \texttt{stage}, \texttt{CVD}, and \texttt{COPD}.
\textit{Model F1-I.II}: frailty model with time-level effect of \texttt{agec} (${\bw}_i$), and hazard-level ($\bx_i$) effects of \texttt{agec}, \texttt{IMD}, \texttt{stage}, \texttt{CVD}, and \texttt{COPD}.  \textit{Models C1-III}, \textit{F1-III}, \textit{C1-IV}, and \textit{F1-IV} represent the corresponding models for Stage III and Stage IV. Results are reported in Table \ref{tab:MLES_strat}. For Stages I-II, the model without frailty is slightly favoured and the estimated frailty variance is small. For Stage III, the frailty model is favoured, and we observe differences in the estimates of the model parameters. Similarly, for Stage IV, the frailty model is favoured, the estimated frailty variance is large, and there is a clear discrepancy in the estimates of the model parameters. Thus, we notice that there are different levels of UIH at each of the tumour stage strata. From Figure \ref{fig:netsurvlung_strat}, we can see that the net survival curves for the models with and without frailty coincide. This indicates that, after stratifying the data by Stage, UIH has a smaller effect on the estimation of net survival, even though there is evidence of UIH for Stages III and IV.
It is worth noticing that omitting important covariates in the models has an effect on the estimation of the baseline hazard parameters and the regression coefficients (see, for instance, the values of the estimates in \ref{tab:MLES} for the models F1, F2, and F3). This is a well known phenomenon associated to omitting important covariates in hazard regression models (see Rossell and Rubio\citep{rossell:2019} for an extensive discussion on the effects of omitting covariates in survival regression models).
      
\begin{table}[ht]
\centering
\begin{tabular}{rrrrrrr}
  \hline
 & C1-I.II & F1-I.II & C1-III & F1-III & C1-IV & F1-IV \\ 
  \hline
$\widehat{\sigma}$    & 2.172 & 2.839 & 0.430 & 1.315 & 0.090 & 0.689 \\ 
$\widehat{\nu}$         & 1.099 & 1.095 & 1.150 & 1.074 & 1.241 & 1.277 \\ 
$\widehat{\gamma}$ & 3.203 & 2.560 & 2.286 & 0.856 & 3.110 & 0.235 \\ 
$\texttt{agec}_t$        & 0.399 & 0.354 & 0.516 & -2.119 & 0.287 & 0.314 \\ 
$\texttt{agec}$           & 0.564 & 0.568 & 0.363 & 0.691 & 0.291 & 0.277 \\ 
$\texttt{IMD}$            & 0.731 & 0.758 & 0.172 & 0.292 & 0.631 & 1.507 \\ 
$\texttt{stage2}$        & 0.770 & 0.836 & --       & --        & --        & --       \\ 
$\texttt{CVD}$           & 0.446 & 0.490 & 0.262 & 0.359 & 0.131 & 0.333 \\ 
$\texttt{COPD}$        & 0.482 & 0.511 & 0.092 & 0.147 & 0.085 & 0.260 \\ 
$\widehat{b}$            &   --     & 0.211 & --        & 0.666 & --        & 3.075 \\ 
AIC                            & \textbf{7817.53} & 7818.49 & 6986.52 & \textbf{6962.63} & 5218.65 & \textbf{5214.64} \\
   \hline
\end{tabular}
\caption{Maximum likelihood estimates for the models fitted on the stratified data by Stage.}
\label{tab:MLES_strat}
\end{table}

\section{Discussion}\label{sec:discussion}
We have developed an extension of frailty models to the relative survival framework, and shown that these models induce a selection of the healthier individuals among survivors of cancer. 
We have shown that the proposed excess hazard frailty models are able to capture unobserved individual heterogeneity, a common challenge in population cancer epidemiology. We have proposed a general family of parametric excess hazard regression models which guarantee identifiability of parameters, and which contains hazard structures of practical interest. The proposed model allows for a tractable implementation of the likelihood function as we can include time-level effects while avoiding the need for numerical integration. Nonetheless, we point out that other hazard structures could be used as well. Although we have focused on the gamma frailty distribution, due to its tractability and interpretability, we have also presented several alternative choices of the frailty distribution that lead to closed-form expressions (see Supplementary material). A thorough comparison of the impact of the choice of the frailty distribution is beyond the aims of this work, but this will be considered in future research.

Our simulations show that the proposed model has good inferential properties. The performance of the proposed approach was evaluated in very challenging scenarios, with a small sample size ($n=500$), substantial amount of censoring (40\%), and with a very complicated generating model (the 4 covariates considered had both an effect on the time scale and on the hazard level). 
In real life applications, one would not expect to see all the covariates with effects on both the time scale and the hazard level. 
As shown in Aim 2 of our simulations, unobserved heterogeneity may have a non-negligible effect if the model fitted to the entire population is used to obtain net survival curves for heterogeneous subgroups. This suggests that discrepancies between net survival curves associated to subgroups of the population using the classical and the frailty models may indicate a non-negligible effect of unobserved heterogeneity, and that a stratified analysis may help reduce this heterogeneity. 

We have shown that, unsurprisingly, the effects and interpretation of UIH that have been reported in the overall survival framework \citep{balan:2020} are also present in the relative survival framework. 
Our work offers an additional perspective, which is the exploration of the impact of UIH on the estimation of marginal quantities (such as net survival). We have shown that the proposed frailty model is able to detect UIH.  
However, it is important to keep the practical limitations of doing so in mind, as the presence of UIH can also be an indication of model inaccuracy or model misspecification \citep{o:2002}.
We have proposed a frailty GH model which allows for including time-varying effects. The use of a general and flexible model thus helps reduce UIH associated to model misspecification. Detecting UIH beyond model misspecification is particularly important when the model fitted to the entire population is also used to estimate marginal quantities associated to subgroups of the population \cite{vaupel:1985}, as the subgroups of interest may have different generating models, or different distributions of the included covariates. 
Thus, in the presence of UIH, it would be useful to compare the results against those obtained with a stratified analysis.
This is an interesting finding for analysts interested in estimating marginal quantities for a subgroup of patients.
Of course, since UIH induces a bias on the estimation of the model parameters (for models without individual frailties), the estimation of conditional quantities (such as excess hazards for specific covariate values) will also be biased. 
We have also shown that UIH induces a selection effect over the follow-up time. This type of effect has been also discussed by Vaupel and Yashin \cite{vaupel:1985}, in the overall survival framework, who found that the estimation of conditional (or individual) quantities can be severely affected by UIH in some scenarios.

We have shown scenarios where a flexible 3-parameter distribution (\textit{e.g.}~PGW) is needed for modelling the shape of the baseline hazard. There exist scenarios where a simpler baseline hazard model (\textit{e.g.}~lognormal) or a simpler hazard structure (such as the AFT or PH) might be favoured by the data. In such cases, a model selection tool (such as AIC or BIC) would help identify the best model for the data (see Rubio et al\cite{rubio:2019S} and Rubio and Drikvandi\cite{rubio:2022} for a discussion).

The real data application presents a clear illustration of the fact that discrepancies in the estimation of the net survival obtained with models with and without frailty terms depend on the distribution of the patients' characteristics. This is in line with the conclusions obtained in Aim 2 of the simulation study. The net survival is a marginal measure, as it is calculated as the average over the observed covariates. Consequently, its estimation naturally depends on the distribution of the covariates. In practice, it is important to first compare the different models in order to detect UIH, followed by an investigation of the effect of the presence of UIH on the estimation of the quantities of interest (such as net survival). Moreover, the real data application indicates that the level of UIH might depend on the observed covariate (\text{e.g.} tumour stage in our illustration). Therefore, extending our approach with a frailty distribution that depends on the observed covariates could be a very interesting research avenue. 

Possible extensions of our work include the use of other parametric baseline hazard distributions, combined with alternative hazard structures (\textit{e.g.}~additive hazard structure). In cancer epidemiology, there are other quantities of interest which are derived from the estimated excess hazard, such as life years lost, among other quantities \citep{belot:2019}. Thus, quantifying the impact of UIH on the estimation of these quantities would be of interest. Since our real data application involves the use of confidential data, we cannot make it publicly available. However, we have created a repository (\url{https://github.com/FJRubio67/IFNS}) which contains an application using ``The Simulacrum'' data set, which is a dataset that contains artificial patient-like cancer data. This repository illustrates the proposed methods and the software provided.

To our knowledge, this is the first study where the impact of UIH on estimated quantities is assessed within the relative survival framework. Given that the net survival is used for international comparisons and to inform policy-making, a careful analysis of factors that may induce bias, such as UIH, on the estimation of this quantity seems appropriate, and the methodology proposed in this paper provides a tool for addressing this problem.


\section*{Financial disclosure}

AB is funded by a Cancer Research UK programme grant (C7923/A29018).

\bibliographystyle{plainnat}
\bibliography{references}  

\end{document}